\documentclass[preprint,12pt]{elsarticle}     % ← uncomment for submission

\usepackage[utf8]{inputenc}
\usepackage[T1]{fontenc}
\usepackage{lmodern}
\usepackage[margin=2.5cm]{geometry}
\usepackage{graphicx}
\usepackage{amsmath,amssymb}
\usepackage{booktabs}
\usepackage{hyperref}
\usepackage{xcolor}
\usepackage{listings}
\usepackage{tikz}
\usetikzlibrary{shapes.geometric, arrows.meta, positioning, calc}
\usepackage{enumitem}
\usepackage[labelfont=bf]{caption}
\usepackage{float}
\usepackage{multirow}
\usepackage{subcaption}
\usepackage{makecell}
\usepackage{tabularx}
\usepackage{lineno}        % CPC requires line numbers for review
\biboptions{numbers,sort&compress}

\definecolor{codebg}{RGB}{245,245,248}
\definecolor{codeframe}{RGB}{200,200,210}
\definecolor{stringcolor}{RGB}{163,21,21}
\definecolor{commentcolor}{RGB}{0,128,0}
\definecolor{keywordcolor}{RGB}{0,0,200}
\definecolor{parnblue}{RGB}{41,98,255}
\definecolor{parngreen}{RGB}{0,150,80}

\hypersetup{
  colorlinks=true,
  linkcolor=blue,
  urlcolor=blue,
  citecolor=blue,
  pdftitle={Parnassus: A GPU-enabled, Python-based Package for Fast Particle Detector Simulation and Reconstruction},
  pdfauthor={Etienne Dreyer, Abdelrahman Elabd, Eilam Gross, Dmitrii Kobylianskii, Benjamin Nachman},
}

\lstdefinestyle{bash}{
  language=bash,
  morekeywords={parnassus,pip,git,cd},
  keywordstyle=\color{keywordcolor}\bfseries,
  commentstyle=\color{commentcolor}\itshape,
  stringstyle=\color{stringcolor},
}

\lstdefinestyle{python}{
  language=Python,
  keywordstyle=\color{keywordcolor}\bfseries,
  commentstyle=\color{commentcolor}\itshape,
  stringstyle=\color{stringcolor},
  morekeywords={Path,Config,GenerationPipeline,JetClusteringPipeline,
    IsolationPipeline,JetClusteringConfig,IsolationConfig,
    RootWriter,HepMC3Generator,self},
}

\lstdefinestyle{yaml}{
  basicstyle=\ttfamily\small,
  backgroundcolor=\color{codebg},
  frame=single,
  rulecolor=\color{codeframe},
  breaklines=true,
  showstringspaces=false,
  commentstyle=\color{commentcolor}\itshape,
  morecomment=[l]{\#},
}

\lstdefinestyle{cmnd}{
  basicstyle=\ttfamily\small,
  backgroundcolor=\color{codebg},
  frame=single,
  rulecolor=\color{codeframe},
  breaklines=true,
  showstringspaces=false,
  commentstyle=\color{commentcolor}\itshape,
  morecomment=[l]{!},
}

\begin{document}

% ── Frontmatter ──────────────────────────────────────────────
\begin{frontmatter}
\title{Parnassus: A GPU-enabled, Python-based Package for Fast Particle Detector Simulation and Reconstruction}

% \author[a]{Etienne Dreyer}
\author[b]{Abdelrahman Elabd}
\author[a]{Eilam Gross}
\author[a]{Dmitrii Kobylianskii}
\author[c,d]{Benjamin Nachman}

\address[a]{Department of Particle Physics \& Astrophysics, Weizmann Institute of Science, Rehovot, Israel}
\address[b]{Department of Physics, University of Washington, Seattle, WA 98195, USA}
\address[c]{Department of Particle Physics and Astrophysics, Stanford University, Stanford, CA 94305, USA}
\address[d]{Fundamental Physics Directorate, SLAC National Accelerator Laboratory, Menlo Park, CA 94025, USA}
% \maketitle

% \bigskip

% ── Abstract ─────────────────────────────────────────────────
\begin{abstract}
We present the public software release of \textsc{Parnassus}, a Python/PyTorch,
GPU-compatible framework for fast detector simulation and reconstruction in
particle and nuclear physics. 
Parnassus provides a user-friendly framework with interchangeable detector
models: neural models can emulate computationally expensive
\textsc{Geant4}-based detector simulation and reconstruction chains, while
parametric models provide PyTorch implementations of selected
\textsc{Delphes}-style detector responses.
This initial release includes two models of the CMS detector:
one based on a flow-matching neural network architecture
and one based on a PyTorch implementation of the Delphes CMS card (parametric bias and smearing).
PyTorch versions of the ATLAS and ALEPH Delphes cards are also available, together
with a flow-matching neural model of the ALEPH detector that extends the framework
to the $e^+e^-$ LEP environment. 
All detector-specific backends share the same
process-agnostic and detector-agnostic API: users select a detector
card---analogous to choosing a detector card in \textsc{Delphes}---and the
same tool can be applied to new physics processes without retraining the
released detector model. 
There are native interfaces to the event generator
\textsc{Pythia} and the event clustering package \textsc{FastJet}.
Unlike previous C++/ROOT-based tools, Parnassus provides GPU-capable 
PyTorch detector-response backends and requires no ROOT installation.  
We describe the installation, command-line and Python API, configuration system, and
demonstrate the framework on Standard Model and BSM processes.
\end{abstract}

% \maketitle

\end{frontmatter}
  
\newpage
% ── CPC Program Summary ─────────────────────────────────────
\bigskip
\begin{center}
\textbf{PROGRAM SUMMARY}
\end{center}

\noindent
\begin{small}
\begin{description}[style=multiline, leftmargin=4.5cm, labelwidth=4.3cm,
  labelsep=0.2cm, font=\normalfont\itshape, itemsep=2pt]

\item[Program title:] Parnassus

\item[Developer's repository link:] \url{https://github.com/parnassus-hep/parnassus}\\
\url{https://parnassus-hep.github.io/parnassus/}

\item[Licensing provisions:] MIT

\item[Programming language:] Python ($\geq$ 3.12)

\item[Nature of problem:]
  Full detector simulation and reconstruction based on \textsc{Geant4} is
  computationally expensive, often dominating the computing budget of
  large-scale particle and nuclear physics experiments and can be 
  computationally prohibitive for individual researchers or large-scale exploratory studies.
  A fast, accurate surrogate is
  needed that maps truth-level (generator-level) particles directly to
  detector-level reconstructed particles, preserving the fidelity of the full
  chain while being orders of magnitude faster.

\item[Solution method:]
  Parnassus offers interchangeable detector backends sharing a common
  API. These backends include both neural network models (based on flow matching) as well as parameterized approaches (\textsc{TorchDelphes}, a re-implementation of selected
  \textsc{Delphes}~\cite{delphes} detector cards). 
  The former has been validated in previous papers~\cite{parnassus_v2,parnassus_lep} 
  and the latter is validated in this work against the C++
  \textsc{Delphes} 3.5.1 cards for the CMS and ATLAS detectors.  
  All backends are followed by physics-based post-processing (e.g., jet clustering via
  \textsc{FastJet}~\cite{fastjet}, lepton isolation calculation). Users select a
  detector card---analogous to a \textsc{Delphes} card---to target a
  specific experiment.

\item[Additional comments including restrictions and unusual features:]
  The framework is implemented in Python/PyTorch and does not require users to build ROOT, C++ Delphes, Pythia8, HepMC, or FastJet by hand. 
  These HEP interfaces are installed as Python package dependencies.
  It has no dependency on ROOT; output to ROOT format is handled via the
  \texttt{uproot}~\cite{uproot} package. 
  The current release ships with CMS, ATLAS, and ALEPH parametric cards and CMS and
  ALEPH neural models;
  additional detector models for other experiments will be added in future releases.

\end{description}
\end{small}

% ══════════════════════════════════════════════════════════════
\section{Introduction}
\label{sec:introduction}

Parnassus is a \textit{Python/PyTorch-based, GPU-compatible} framework for
\textit{fast detector simulation and reconstruction} in particle physics.
It provides an alternative path to computationally expensive full
\textsc{Geant4}~\cite{geant4} simulation and reconstruction chains with
generative models that map truth-level (generator-level) particles directly to
detector-level reconstructed particles,
bypassing intermediate steps such as hit simulation, digitisation, and track
reconstruction. 
The framework is both \textit{process-agnostic} and
\textit{detector-agnostic}: users select a detector card to target a specific
experiment, and the same tool can be applied to new physics processes by
providing different input truth events, subject to the validated phase space of
the selected detector model.

Parnassus includes a number of interchangeable \textit{generator backends} sharing a
common API; these backends are of two types:
\begin{itemize}
  \item \textit{Neural backends} based on conditional flow
    matching~\cite{lipman2023flow,liu2023flow} trained on full simulation and
    reconstruction data, providing high-fidelity learned detector response.
    The models are conditioned on
    truth-level particle properties, making the generation inherently
    process-agnostic without requiring retraining for new physics scenarios.
    Process-level generalization, including out-of-distribution studies for
    the released CMS-style neural model, was studied in Ref.~\cite{parnassus_v2}.
  \item A \textit{parametric backend}, \textsc{TorchDelphes}, which
    re-implements the \textsc{Delphes}~\cite{delphes} fast-simulation chain
    (particle propagation, tracking efficiency, momentum smearing, calorimeter
    response, energy-flow merging) in pure PyTorch. The implementation is
    validated against C++~\textsc{Delphes} 3.5.1 in~\ref{section:torch_delphes_validation} for both CMS and ATLAS
    detector cards and provides a GPU-native drop-in alternative to running
    C++ \textsc{Delphes}.
\end{itemize}
A list of models available with the current release can be found in~\autoref{tab:released-backends}.

Unlike traditional fast simulation tools such as
\textsc{Delphes}~\cite{delphes}, which are written in C++, require the ROOT
framework~\cite{root}, and run only on CPU, the detector-response backends in
Parnassus are implemented in Python/PyTorch~\cite{pytorch} and can run on GPU, enabling seamless integration into modern Python-based analysis workflows
and significant acceleration on GPU hardware. 
Parnassus has no dependency on ROOT itself; ROOT I/O is supported via
\texttt{uproot}~\cite{uproot} for writing output files.

The physics methodology underlying Parnassus has been described in detail in
Refs.~\cite{parnassus_v1,parnassus_v2}, where we demonstrated that the approach
reproduces full CMS simulation and reconstruction across a wide range of
Standard Model processes. 
In this document, we focus on the \textit{public software release} --- its installation, usage, and API --- to enable the broader community to adopt neural fast simulation and reconstruction in their workflows. 

% ══════════════════════════════════════════════════════════════
\section{Architecture overview}
\label{sec:architecture}

\begin{figure}[htbp]
\centering
\includegraphics[width=0.5\textwidth]{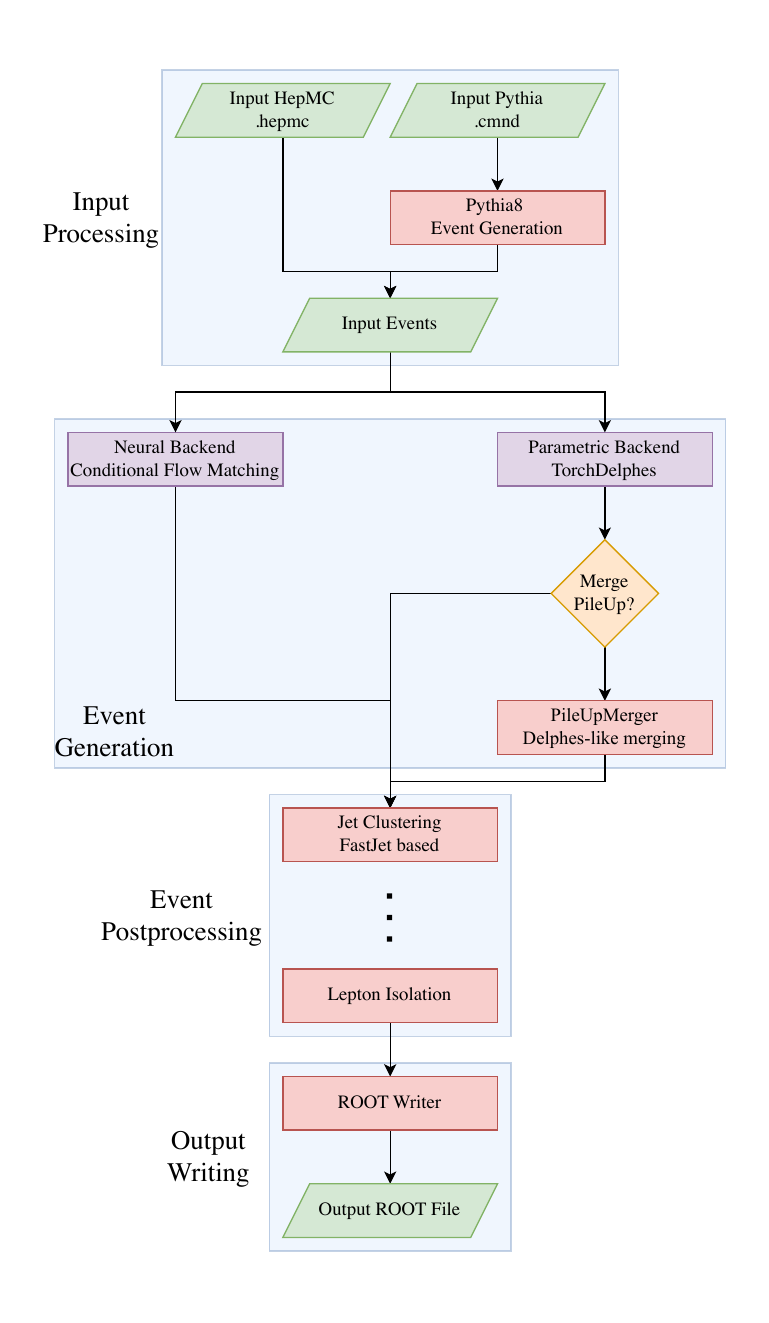}
\caption{Parnassus generation pipeline. Truth-level events are routed through
  either the neural backend (conditional flow matching) or the parametric
  backend (\textsc{TorchDelphes}), both producing particle-flow objects.
  Physics-based event post-processing (e.g. jet clustering) computes derived quantities and
  writes the final output.}
\label{fig:architecture}
\end{figure}

Parnassus passes truth-level events through a detector model followed by
physics-based post-processing. 
The detector model can be either a neural
backend or the \textsc{TorchDelphes} parametric backend; both share the same
input and output interface so that downstream post-processing is unchanged.
The pipeline is illustrated in~\autoref{fig:architecture} and can be split into
four parts: input processing, event generation, event post-processing, and
output writing.

\begin{table}[htbp]
\centering
\caption{Detector backends included in the current Parnassus release.}
\label{tab:released-backends}
\scriptsize
\begin{tabular}{@{}llll l@{}}
\toprule
\textbf{Backend} & \textbf{Detector card} & \textbf{Validation} &
\textbf{Particle limit} & \textbf{Pile-up treatment} \\
\midrule
Neural flow model & CMS 2011 & Ref.~\cite{parnassus_v2} &
400 Truth/PFlow particles & \makecell[l]{Fixed by training sample;\\implicit PFlow PU} \\
Neural flow model & ALEPH & Ref.~\cite{parnassus_lep} &
128 Truth/PFlow particles & N/A \\
\textsc{TorchDelphes} & CMS & \makecell[l]{This work;\\C++ \textsc{Delphes} 3.5.1} &
No fixed model limit & MinBias merging \\
\textsc{TorchDelphes} & ATLAS & \makecell[l]{This work;\\C++ \textsc{Delphes} 3.5.1} &
No fixed model limit & MinBias merging \\
\textsc{TorchDelphes} & ALEPH & \makecell[l]{This work;\\C++ \textsc{Delphes} 3.5.1} &
No fixed model limit & N/A \\
\bottomrule
\end{tabular}
\end{table}

\subsection{Input processing}
Parnassus takes as input either \textsc{HepMC}~\cite{buckleyHepMC3EventRecord2021}
event files or \textsc{Pythia8}~\cite{bierlich_comprehensive_2022}
\texttt{.cmnd} steering cards. 
For the latter, it runs the bundled Pythia8
interface to obtain truth particles, which are then used to generate PFlow
events.
\subsection{Event Generation}
For event generation, Parnassus supports two possible backends: neural and parametric. Each of them has a pre-defined set of ``cards'', similar to Delphes cards, that represent the different detector conditions.
\begin{itemize}
    \item The \textbf{neural backend} is a continuous-time generative model trained with
    conditional flow matching~\cite{lipman2023flow,liu2023flow} and sampled via
    Euler integration of the learned vector field. 
    The model is conditioned on truth-level particle information, 
    making the generation process inherently
    dependent on the input physics process without being tied to any specific one.
    The number of integration steps is configurable at runtime, 
    allowing users to trade generation speed for quality.
    The current release provides the CMS\footnote{An ATLAS neural flow model is under development~\cite{ATLAS_SIMU_2026_09}.} and ALEPH detector cards with a newly
    trained models using a similar architecture, model size, and
    training-sample definition as Refs.~\cite{parnassus_v2, parnassus_lep}. It is not a
    re-export of the checkpoint used in those studies.  
    \item The \textbf{parametric backend}, \textsc{TorchDelphes}, is a pure-PyTorch
    re-implementation of the \textsc{Delphes}~\cite{delphes} fast-simulation chain 
    running natively on GPU.
    It models particle propagation through the magnetic field, tracking
    efficiency, momentum smearing, calorimeter response, and energy-flow merging,
    with per-module parameters loaded from the selected detector card. 
    The current release provides CMS, ATLAS and ALEPH cards ported from their respective
    \textsc{Delphes} Tcl cards and validated against C++~\textsc{Delphes}~3.5.1.
    When configured with a pile-up block, the parametric path first merges
    minimum-bias interactions with the hard-scatter event using a
    Delphes-compatible \texttt{PileUpMerger} before applying the detector card.
\end{itemize}
Both backends expose the same event-level output scheme (particle-flow kinematics,
vertex positions, particle classification, and, where supported, track impact
parameters), so downstream post-processing is identical.

Neural backends support the fixed number of constituents used during training;
for example, the CMS 2011 model supports up to 400 truth or PFlow particles 
while ALEPH model supports up to 128.
The parametric backend does not have this model-imposed limitation.

Since training datasets with pileup truth particles are not publicly available, the models have to learn unconditional modeling of pileup.  As noted above, this is treated with an overlay method in the parametric backend and manifests as `fake' (in the sense of not corresponding to a truth-level particle) particle production.  Unlike the parametric model, the neural model can readily add additional particles.  The model would be more precise if conditioned on truth pileup, as it would then only have to model the detector response and not also the underlying spectrum.  Research is ongoing into the best ways to improve pileup modeling along this direction.

\subsection{Backend selection}
\label{subsec:backend-selection}

The detector backend is selected entirely through the configuration file. 
The same input file, post-processing configuration, and output schema can therefore
be used with either the neural backend or the parametric \textsc{TorchDelphes}
backend. 
Switching from the neural CMS model to the parametric CMS card only
changes the generator block:

\begin{lstlisting}[style=yaml,caption={Alternative generator blocks for changing
the detector backend while leaving the rest of the workflow unchanged.},
label=lst:swap-backends]
# Neural CMS model
generator:
  type: "neural"
  name: "cms_2011_flow_v00"
  num_steps: 50
  batch_size: 2000
  device: "cuda"

# Parametric CMS card
generator:
  type: "parametric"
  name: "cms"
  seed: 42
\end{lstlisting}

The resulting ROOT files follow the same collection naming convention, so the
same downstream analysis code can read neural and parametric outputs. Switching
from the CMS to the ATLAS parametric card requires changing
\texttt{generator.name} from \texttt{"cms"} to \texttt{"atlas"}.

\subsection{Event Post-processing}
After event generation, optional post-processing is applied.
For computational efficiency, the user can run post-processing in parallel by specifying the number of processes in the configuration file.
The current software release supports three post-processing pipelines useful for analysis purposes: particle filtering, jet clustering, and charged lepton isolation calculation.
\begin{itemize}
    \item Particle filtering drops particles from a target collection in place
    according to declarative per-field conditions (e.g. $p_T$, $\eta$, or PDG-ID
    cuts), combined with logical \texttt{all} (AND) or \texttt{any} (OR) semantics.
    It can be applied to the Truth, PFlow, lepton, or any generator-specific
    collection (e.g. \texttt{Track}, \texttt{Tower}). Because it mutates the
    collection in place, a filtering pipeline is declared \emph{before} clustering
    or isolation so that those stages operate on the surviving particles only.
    By default the cached scalar event features ($H_T$ and $\vec{E}_T^{\mathrm{miss}}$)
    are recomputed from the survivors to remain consistent with the cut collection.
    \item Jet clustering is done via the FastJet~\cite{fastjet} library, with four supported clustering algorithms: $e^+e^-$ generalized kT, generalized kT, anti-kT, and Cambridge/Aachen algorithms.
    It can be performed on both Truth and PFlow particle collections with customizable cuts on jet $p_T$ and number of constituents.
    \item The isolation of individual $e$ or $\mu$ leptons is measured relative
    to their transverse momentum $p_T^\ell$ by summing over charged and
    neutral particles in a cone
    $\Delta R = \sqrt{\Delta \eta^2 + \Delta \phi^2} < 0.4$ around the lepton
    direction:
    \begin{equation}
    R^\ell_{\mathrm{iso}} =
    \frac{
        \sum_{\Delta R<0.4} p_T^{\mathrm{charged}}
        +
        \sum_{\Delta R<0.4} p_T^{\mathrm{neutral}}
        +
        \sum_{\Delta R<0.4} p_T^\gamma
    }{p_T^\ell},
    \end{equation}
    where the sums exclude the lepton itself.
    The last term excludes photons that are candidates for final-state radiation (FSR) from the lepton.
\end{itemize} 

\subsection{Output Writing}
Output writing is performed via the \texttt{uproot}~\cite{uproot} library, allowing ROOT I/O without installing ROOT. 
A short summary of output collections is given in~\autoref{tab:outputs}.
\begin{table}[htbp]
\centering
\caption{Output quantities produced by Parnassus for each event.}
\label{tab:outputs}
\small
\begin{tabular}{@{}ll@{}}
\toprule
\textbf{Output} & \textbf{Description} \\
\midrule
Reconstructed particles & Full kinematics ($p_\mathrm{T}$, $\eta$, $\phi$), vertex coordinates
  ($v_x$, $v_y$, $v_z$), Class\,ID, PDG\,ID \\
\makecell[l]{Impact parameters\\(if model supports)} & $d_0$, $z_0$ and their errors for charged particles \\
Jets & Jet kinematics ($p_T$, $\eta$, $\phi$), $D_2$ and $C_2$ substructure variables. \\
Lepton isolation & Isolation scores and $p_\mathrm{T}$ sums in configurable
  $\Delta R$ cones, etc. \\
Event-level quantities & Event number, $H_\mathrm{T}$, $\vec{E}_\mathrm{T}^\mathrm{miss}$
  values for Truth and PFlow collections. \\
\bottomrule
\end{tabular}
\end{table}\\
The exact set of written branches is determined by the selected backend and
post-processing pipelines. 
% ══════════════════════════════════════════════════════════════

\section{Computational performance}
\label{sec:performance}

The wall-clock performance of \textsc{Parnassus} depends on the detector backend, event multiplicity, batch size, hardware, and whether additional post-processing steps are enabled. For the benchmark in \autoref{tab:performance}, we first generated 10,000 $H\to ZZ\to 4\ell$ events with Pythia8 and stored them in HepMC3 format. This HepMC3 file was then used as a common input for all workflows. 

All optional post-processing pipelines, such as jet clustering and lepton isolation, were disabled. The reported throughput therefore measures the core event-processing workflow: input reading, detector-response generation, and ROOT output writing. For \textsc{TorchDelphes} and \textsc{Delphes}, we enabled pile-up merging to match the neural backend.  The neural backend was not optimized for speed and could be accelerated with additional studies.  The current rate will likely not be the limiting factor for most applications.

\begin{table}[htbp]
\centering
\small
\begin{tabular}{@{}lllllll@{}}
\toprule
\textbf{Backend} & \textbf{Card} & \textbf{Device} &
\textbf{Batch size} & \textbf{Steps} & \multicolumn{2}{l}{\textbf{Throughput}}\\
 & & & & & \textbf{(events/s)} & \textbf{(s/events)} \\
\midrule
Neural flow & CMS 2011 & GPU & 1024 events & 50 & 17.5 & 0.0571 \\
\textsc{TorchDelphes} & CMS & GPU & 1024 particles & -- & 86.9 & 0.0115 \\
\textsc{Delphes} 3.5.1 & CMS & CPU & -- & -- & 80.6 & 0.0124 \\
\bottomrule
\end{tabular}
\caption{Event-processing throughput for the CMS backends. All measurements use
10,000 pre-generated $H\to ZZ\to4\ell$ HepMC3 events as input, with
post-processing pipelines disabled. The GPU measurements were performed on an
NVIDIA RTX A6000 Ada Generation GPU utilizing Tensor Cores, while the CPU measurement used an Intel Xeon Gold
6530 CPU.}
\label{tab:performance}
\end{table}

% ══════════════════════════════════════════════════════════════
\section{Demonstration: SM and BSM processes}
\label{sec:demo}

The neural backend is \textbf{process-agnostic} --- it learns the detector
response as a function of truth-level particle properties, not the physics
process itself --- and the parametric (\textsc{TorchDelphes}) backend is
process-agnostic by construction. We have previously
demonstrated~\cite{parnassus_v1,parnassus_v2} that Parnassus reproduces full
simulation and reconstruction across a wide range of SM processes.
Here we show two representative examples to illustrate the framework in action: a
Standard Model process ($H \to ZZ \to 4\ell$) and a BSM exotic Higgs decay
($H \to aa \to gg\mu\mu$). 
Both examples and the distributions shown in this
section use the neural CMS backend. 
These examples demonstrate the public
workflow and are not intended as a new validation against full detector
simulation; validation of the neural model is described in
Refs.~\cite{parnassus_v1,parnassus_v2}.

\subsection{Standard Model: $H \to ZZ \to 4\ell$}

For the SM example, we generate Higgs events on the fly with
\textsc{Pythia8}~\cite{bierlich_comprehensive_2022} using the
\texttt{h4lep.cmnd} file shown in~\autoref{code:h4lep}.
The Higgs is produced via gluon fusion, and we require decays to two $Z$ bosons, each decaying into two leptons.

\begin{lstlisting}[style=bash]
parnassus run \
  -c neural_config.yaml \
  -i h4lep.cmnd \
  -ne 50_000 \
  -o h4l_output.root
\end{lstlisting}

\begin{figure}[htbp]
\centering
\includegraphics[width=\textwidth]{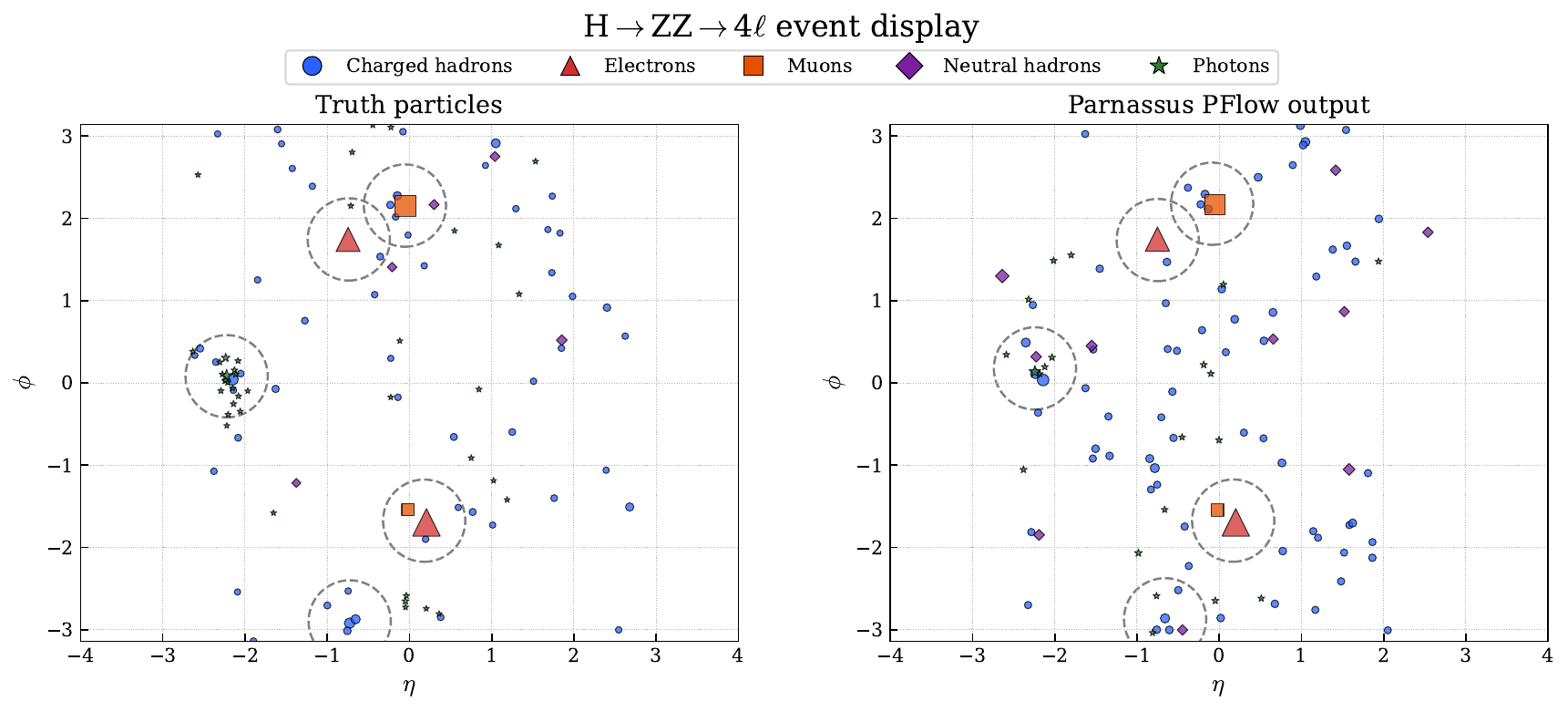}
\caption{Example event display for $H \to ZZ \to 4\ell$: truth particles (left) and
  Parnassus PFlow output (right) in the $\eta$--$\phi$ plane. Marker size
  scales with $p_\mathrm{T}$; dashed circles indicate anti-$k_t$ jets
  ($R = 0.5$). Particle classes: charged hadrons (blue circles), electrons
  (red triangles), muons (orange squares), neutral hadrons (purple diamonds),
  photons (green stars). }
\label{fig:display_h4l}
\end{figure}

\autoref{fig:display_h4l} shows a representative $H \to 4\ell$ event
in the $\eta$--$\phi$ plane, illustrating how Parnassus transforms
truth-level particles into detector-level PFlow objects.

\subsection{BSM: $H \to aa \to gg\mu\mu$}

\begin{figure}[htbp]
\centering
\includegraphics[width=\textwidth]{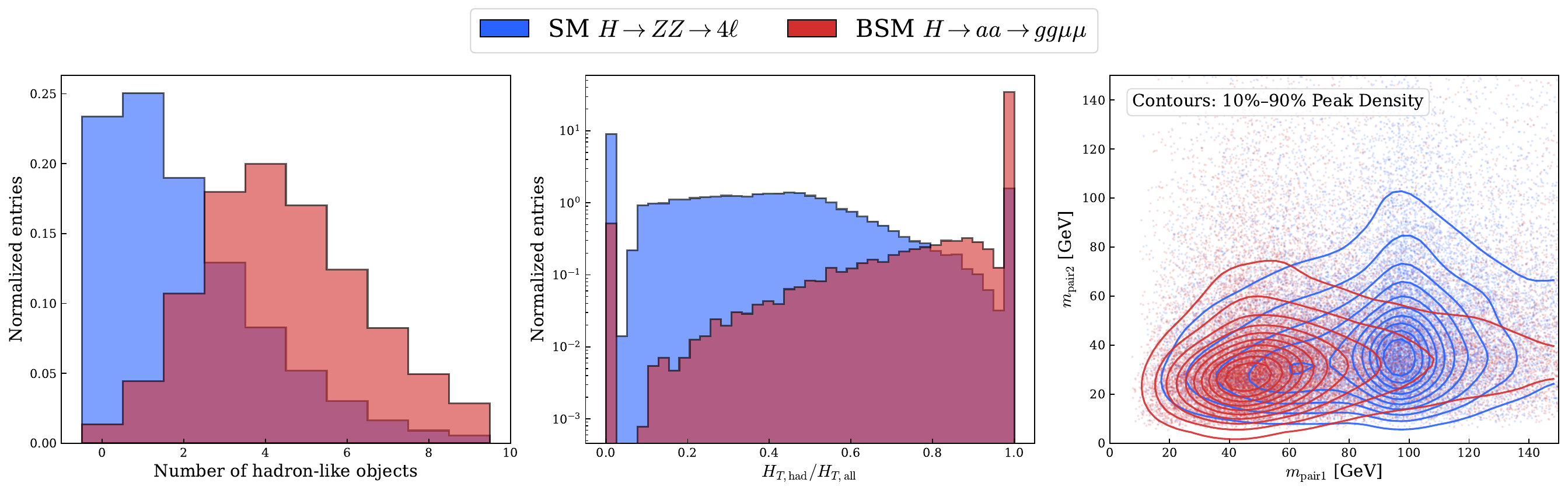}
\caption{PFlow-level comparison of SM
  ($H{\to}ZZ^{(*)}{\to}4\ell$, 50,000 events) and BSM
  ($H\to aa$, $a\to gg$ or $\mu^+\mu^-$, 50,000 events) Higgs processes. All PFlow
  candidates, including leptons, are clustered into anti-$k_t$ objects with
  $R=0.5$ and $p_\mathrm{T}>10$\,GeV. The distributions illustrate differences
  in hadron-like activity and visible resonance structure between the two
  processes.}
\label{fig:process_comparison}
\end{figure}

For the BSM example, we generate exotic Higgs decays on-the-fly with
\textsc{Pythia8} with \texttt{haa\_ggmumu.cmnd} shown in~\autoref{code:bsm}.
The Higgs decays to a pair of light pseudoscalars $a$
($m_a = 20$\,GeV), each decaying to either $gg$ or $\mu^+\mu^-$, producing
a mixed final state of jets and muon pairs:

\begin{lstlisting}[style=bash]
parnassus run \
  -c neural_config.yaml \
  -i haa_ggmumu.cmnd \
  -ne 50_000 \
  -o haa_ggmumu_output.root
\end{lstlisting}

\noindent

\autoref{fig:process_comparison} compares PFlow-level observables for the SM
and BSM Higgs samples after clustering all PFlow candidates, including leptons,
into anti-$k_t$ objects with $R=0.5$ and $p_\mathrm{T}>10$\,GeV. 
Since no lepton cleaning is applied, isolated leptons are reconstructed as narrow
anti-$k_t$ objects. 
The resulting object collection therefore captures the full
visible final-state topology rather than only hadronic jet activity.

The two processes lead to qualitatively different visible-object structure. 
In the BSM sample, $H\to aa\to gg\mu\mu$, the $a\to gg$ decay produces genuine
hadronic activity, while the muons form lepton-like objects. 
In contrast, the SM sample, $H\to ZZ^{(*)}\to4\ell$, is dominated by lepton-like objects, with
hadronic activity arising mainly from initial-state radiation and the
underlying event. 
This difference is reflected in the hadron-like object multiplicity and in the fraction of visible transverse momentum carried by hadron-like objects.

We further compare pair-mass observables constructed from the four leading
anti-$k_t$ objects. 
In the BSM process, the pair masses are sensitive to the two intermediate pseudoscalar resonances and are therefore expected to show
structure near $m_a=20$\,GeV for correctly paired objects. 
In the SM process, because leptons are included in the clustering, the same observable is
sensitive to the $ZZ^{(*)}\to4\ell$ topology: one object pair can reconstruct a
mass near the $Z$-boson mass, while the second pair is typically off shell.
These distributions demonstrate that the same Parnassus workflow can be applied 
to qualitatively different final states and produces reconstructed-level observables 
that reflect the expected visible topologies.

% ══════════════════════════════════════════════════════════════

% ══════════════════════════════════════════════════════════════
\section{Conclusion}
\label{sec:conclusion}

We have presented the public release of Parnassus, a Python/PyTorch,
GPU-compatible framework for fast simulation and reconstruction in high
energy physics. 
Unlike traditional C++/ROOT-based tools such as
\textsc{Delphes}~\cite{delphes}, Parnassus is used through a Python/PyTorch
interface, supports GPU acceleration, and requires no ROOT or C++
\textsc{Delphes} installation --- ROOT output is optionally handled via
\texttt{uproot}.
The software provides a simple command-line and Python API for generating
detector-level particle-flow objects from truth-level input, with full
post-processing including jet clustering and lepton isolation.

The current release includes CMS and ALEPH neural detector models and CMS, ATLAS,
and ALEPH \textsc{TorchDelphes} detector cards, all exposed through the same
configuration and output conventions.

The neural model validation and process-generalization studies are documented
in Refs.~\cite{parnassus_v1,parnassus_v2}; this software release makes those
models usable in a public workflow and adds the GPU-native parametric
\textsc{TorchDelphes} backend. Future releases will extend the library of
detector cards and broaden the set of validated detector configurations.

Parnassus is open-source under the MIT License; its source code and 
detailed documentation can be found at \url{https://github.com/parnassus-hep/parnassus} 
and \url{https://parnassus-hep.github.io/parnassus/}, respectively.

\section*{Acknowledgements}
This work was supported in part by the U.S. Department of Energy (DOE) under Contract No. DE-AC02-76SF00515 and by the CompHEP Center for Computational Excellence, and by the U.S. National Science Foundation under Grant No. OAC-2417682. DK and EG are supported by the Minerva Grant 715027, and the Knell Family Institute for Artificial Intelligence.  AE is supported by a DOE CompHEP Western Advanced Training for Computational High-Energy Physics (WATCHEP) Fellowship, under DOE Award DE-SC-0023527. This research used resources of the National Energy Research Scientific Computing Center (NERSC), a DOE Office of Science User Facility supported by the Office of Science of the U.S. Department of Energy under Contract No. DE-AC02-05CH11231, under NERSC award HEP-ERCAP0035546.

% ══════════════════════════════════════════════════════════════
%% CPC uses numerical citations with elsarticle-num.bst.
%% For submission, replace the thebibliography block with:
%%   \bibliographystyle{elsarticle-num}
%%   \bibliography{parnassus}
%% and provide a parnassus.bib file.

\appendix
\section{Program reference: installation and quick start}
\label{sec:quickstart}

This appendix summarizes the user-facing program interface. It is intended as a
compact reference for installation, backend selection, command-line execution,
programmatic execution, configuration, output structure, and the input cards
used in Sec.~\ref{sec:demo}.

\subsection{Installation}
Installing the Parnassus package requires Python 3.12 or higher.
\begin{lstlisting}[style=bash,caption={Installation commands.}]
# Clone the repository
git clone https://github.com/parnassus-hep/parnassus.git
cd parnassus

# Install with pip (requires Python 3.12)
pip install .

# For development
uv sync --all-extras
\end{lstlisting}

The pretrained CMS neural checkpoints are stored inside the public repository,
so cloning the repository also downloads the released model weights.
The Python package installation installs the required HEP interfaces, including
\textsc{Pythia8}, HepMC, \textsc{FastJet}, PyTorch, and \texttt{uproot}; users
do not need to build ROOT, C++ \textsc{Delphes}, \textsc{Pythia8}, HepMC, or
\textsc{FastJet} manually for the standard workflows.

\subsection{Initialize a configuration file}

\begin{lstlisting}[style=bash]
parnassus init .
\end{lstlisting}

This copies two default configuration files to the current directory:
\texttt{neural\_config.yaml} (using the \texttt{cms\_2011\_flow\_v00} neural
detector model) and \texttt{parametric\_config.yaml} (using the CMS
\textsc{TorchDelphes} card). 
Both configurations include anti-$k_t$ jet
clustering and lepton isolation pipelines. 
To target a different detector or
to switch backends, change the \texttt{generator.type} and
\texttt{generator.name} fields in the YAML configuration. 

The released neural CMS model was trained on 2011 CMS pile-up conditions. 
It is conditioned on hard-scatter truth particles rather than explicit truth pile-up
particles; pile-up-like PFlow particles are therefore generated implicitly by
the model and are not configured through a pile-up block. 
The neural CMS model supports up to 400 truth particles after the model selection cuts.
The two generator backends are configured as shown in~\autoref{lst:generator-backends}.

\begin{lstlisting}[style=yaml,caption={Switching between the neural and
parametric (\textsc{TorchDelphes}) generator backends.},
label=lst:generator-backends]
# --- Neural backend (neural_config.yaml) ---
generator:
  type: "neural"
  name: "cms_2011_flow_v00"  # trained CMS model
  num_steps: 50              # Euler integration steps
  batch_size: 2000
  device: "cpu"              # "cpu", "cuda", or "mps"

# --- Parametric backend (parametric_config.yaml) ---
generator:
  type: "parametric"
  name: "cms"                # "cms" or "atlas" detector card
  seed: 42                   # optional, for reproducibility
  # pileup:                  # optional Delphes-style pile-up
  #   file_path: "MinBias.pileup"
  #   mean_pileup: 50
\end{lstlisting}

\subsection{Run with a HepMC input file}

\begin{lstlisting}[style=bash,caption={Basic command-line invocation.}]
parnassus run \
  -c neural_config.yaml \
  -i events.hepmc \
  -ne 1000 \
  -bs 2000 \
  -o output.root
\end{lstlisting}

\noindent
The available command-line flags are listed in~\autoref{tab:cli}.

\begin{table}[htbp]
\centering
\caption{Command-line flags for \texttt{parnassus run}.}
\label{tab:cli}
\begin{tabular}{@{}lll@{}}
\toprule
\textbf{Flag} & \textbf{Long form} & \textbf{Description} \\
\midrule
\texttt{-c} & \texttt{-{}-config}      & Path to YAML configuration file \\
\texttt{-i} & \texttt{-{}-input\_path}  & Input HepMC or Pythia \texttt{.cmnd} file \\
\texttt{-o} & \texttt{-{}-output\_path} & Output ROOT file path \\
\texttt{-ne} & \texttt{-{}-num\_events} & Number of events to process \\
\texttt{-bs} & \texttt{-{}-batch\_size} & Batch size for processing \\
\texttt{-n} & \texttt{-{}-num\_steps}   & Neural-backend Euler integration steps (default: 50) \\
 & \texttt{-{}-random\_seed} & Random seed for parametric generators (reproducibility) \\
\bottomrule
\end{tabular}
\end{table}

\subsection{Run with Pythia8 on-the-fly generation}

To generate truth-level events on-the-fly with \textsc{Pythia8}, create a
\texttt{.cmnd} steering card and pass it as the input file. Parnassus detects
\texttt{.cmnd} files automatically:

\begin{lstlisting}[style=bash]
parnassus run \
  -c neural_config.yaml \
  -i h4lep.cmnd \
  -ne 5000 \
  -bs 2000 \
  -o h4lep_fastsim.root
\end{lstlisting}

When running the neural backend aimed at the CMS detector, 
it is important to match the CMS convention for stable particles ($c\tau > 10$ mm). 
This can be done by setting the following block in the Pythia8 configuration:
\begin{lstlisting}[style=cmnd]
ParticleDecays:limitTau0 = on
ParticleDecays:tau0Max = 10          ! mm/c
\end{lstlisting}
These settings make particles with $c\tau > 10$ mm stable in the Pythia event
record, matching the stable-particle convention used to train the CMS neural
backend. 
Inputs generated with a different convention may fall outside the model's expected domain.

\subsection{Input and output overview}

The two user-facing input modes are HepMC event files and Pythia8 steering
cards. HepMC input is appropriate when truth events have already been generated
externally. Pythia8 \texttt{.cmnd} input is appropriate when Parnassus should
generate truth events on the fly before detector simulation.

\begin{table}[htbp]
\centering
\caption{User-facing input modes.}
\label{tab:input-modes}
\begin{tabular}{@{}lll@{}}
\toprule
\textbf{Input} & \textbf{Extension} & \textbf{Use case} \\
\midrule
HepMC3 event record & \texttt{.hepmc} & Read pre-generated truth-level events \\
Pythia8 steering card & \texttt{.cmnd} & Generate truth-level events on the fly \\
\bottomrule
\end{tabular}
\end{table}

Parnassus writes one ROOT file using \texttt{uproot}; no ROOT installation is
required. The output file contains a single tree named \texttt{Parnassus}.
Branches use flat dot-separated names of the form
\texttt{Collection.Field}. The main collections are listed in~\autoref{tab:output-overview}.

\begin{table}[htbp]
\centering
\caption{Main output collections. Jet collections are created from the
configured clustering pipelines and use the pipeline name as the collection
name.}
\label{tab:output-overview}
\begin{tabular}{@{}llll@{}}
\toprule
\textbf{Collection} & \textbf{Neural} & \textbf{Parametric} & \textbf{Description} \\
\midrule
\texttt{Truth} & yes & yes & Truth-level stable input particles \\
\texttt{PFlow} & yes & yes & Detector-level particle-flow candidates \\
\texttt{Track} & no & yes & Reconstructed charged tracks \\
\texttt{Tower} & no & yes & Calorimeter tower deposits \\
\texttt{Electrons} & yes & yes & Electron kinematics and isolation fields \\
\texttt{Muons} & yes & yes & Muon kinematics and isolation fields \\
\texttt{Event} & yes & yes & Event number, $H_T$ and MET components \\
\texttt{<JetName>} & optional & optional & One collection per clustering pipeline \\
\bottomrule
\end{tabular}
\end{table}

\section{Python API}
\label{sec:api}

For programmatic usage, Parnassus provides a Python API. 
The command-line interface runs the same sequence automatically;~\autoref{lst:api} shows the
main loop explicitly, including generation, post-processing, inspection, and
output writing.

\begin{lstlisting}[style=python,caption={Programmatic usage of the generation pipeline.},label=lst:api]
from pathlib import Path
from parnassus.configs import Config
from parnassus.pipelines import (
    GenerationPipeline,
    JetClusteringPipeline,
    IsolationPipeline,
)
from parnassus.configs.pipeline import (
    JetClusteringConfig,
    IsolationConfig,
)
from parnassus.writers import RootWriter

# Load configuration
config = Config.from_yaml("my_config.yaml")

# Define output path
config.writer_config.file_path = Path("test_output.root")

# Get accessor store for output writing (e.g. to define new accessors based on generated data)
accessor_store = config.writer_config.accessor_store

# Override settings programmatically
config.dataset_config.file_path = (
    Path("events.hepmc").absolute()
)
config.dataset_config.num_events = 50

# Run generation
pipeline = GenerationPipeline(config)
gen_events, accessors_dict = pipeline.run()

# Update accessor store with any new accessors created during generation
accessor_store.update_from_dict(accessors_dict)

# Post-processing
for pipeline_config in config.pipeline_configs:
    if isinstance(pipeline_config, JetClusteringConfig):
        jet_pipeline = JetClusteringPipeline(pipeline_config)
        jet_pipeline.process(gen_events)
        accessor_store.update_from_dict(jet_pipeline.get_accessors())
    elif isinstance(pipeline_config, IsolationConfig):
        iso_pipeline = IsolationPipeline(pipeline_config)
        iso_pipeline.process(gen_events)
        accessor_store.update_from_dict(iso_pipeline.get_accessors())

# Inspect generated data
for event in gen_events[:5]:
    print(f"Event {event.event_number}:")
    n_truth = len(event.truth_particles.pt)
    n_pflow = len(event.pflow_particles.pt)
    print(f"  Truth particles: {n_truth}")
    print(f"  PFlow particles: {n_pflow}")
    print(f"  Truth HT = {event.truth_ht:.1f} GeV")
    print(f"  PFlow HT = {event.pflow_ht:.1f} GeV")

# Print what will be written to the output file
print(accessor_store)

# Write to ROOT
writer = RootWriter(config.writer_config)
writer.write(gen_events)
\end{lstlisting}

\subsection{Reading output files}

\begin{lstlisting}[style=python,caption={Reading the output ROOT file with \texttt{uproot}.}]
import uproot
import numpy as np

f = uproot.open("output.root")
tree = f["Parnassus"]

# Access pflow jet pT
jet_pt = tree["PFlowJetsAntiKt05.PT"].array()

# Access electron isolation
e_iso = tree["Electrons.IsolationVar"].array()

# Access impact parameters
d0 = tree["PFlow.D0"].array()
z0 = tree["PFlow.Z0"].array()
\end{lstlisting}

% ══════════════════════════════════════════════════════════════
\section{Configuration reference}
\label{sec:config}

Parnassus uses a single YAML configuration file. 
The configuration has four top-level blocks: 
\texttt{dataset}, \texttt{generator}, \texttt{pipelines}, and
\texttt{output}. 
The \texttt{generator} block selects the detector backend, so
backend switching is a configuration change rather than a change to the
analysis code.

\begin{lstlisting}[style=yaml,caption={Annotated default configuration
(\texttt{neural\_config.yaml})},label=lst:config]
# -- Dataset -------------------------------------------
dataset:
  file_path: ""              # Path to .hepmc or .cmnd file
  num_events: 1000           # Number of events to process
  entry_start: 0             # Starting event index (HepMC)

# -- Generator (choose one backend) --------------------
# Option A: Neural backend (conditional flow matching)
# CMS 2011 model
generator:
  type: "neural"             # Select the neural backend
  name: "cms_2011_flow_v00"  # Detector model (like a Delphes card)
  num_steps: 50              # Number of flow-matching integration steps
  batch_size: 2000           # Events per batch
  device: "cpu"              # "cpu", "cuda", or "mps"
  
# Option B: Neural backend (conditional flow matching)
# ALEPH model
# generator:
#   type: "neural"             # Select the neural backend
#   name: "aleph_flow_v00"     # Detector model (like a Delphes card)
#   num_steps: 50              # Number of flow-matching integration steps
#   batch_size: 2000           # Events per batch
#   device: "cpu"              # "cpu", "cuda", or "mps"

# Option C: Parametric backend (TorchDelphes)
# generator:
#   type: "parametric"       # Select the TorchDelphes backend
#   name: "cms"              # Detector card: "cms" or "atlas"
#   seed: 42                 # Random seed for stochastic smearing

# -- Post-processing Pipelines ------------------------
pipelines:
  # ParticleFilter:           # Optional: drop particles before clustering/isolation
  #   type: "filter"
  #   collection: truth       # truth | pflow | electrons | muons | <collection key>
  #   combine: all            # all (AND) | any (OR)
  #   conditions:
  #     - {field: pt,  op: ">",  value: 0.5}
  #     - {field: eta, op: "<=", value: 3.0, abs: true}
  TruthJetsAntiKt05:
    type: "cluster"
    collection: truth
    batch_size: 2000          # Optional post-processing batch size
    num_processes: 2          # Optional number of worker processes
    dr: 0.5
    algorithm: antikt
    pt_min: 10
    nconst_min: 2
  TruthJetsAntiKt08:
    type: "cluster"
    collection: truth
    dr: 0.8
    algorithm: antikt
    pt_min: 10
    nconst_min: 2
  PFlowJetsAntiKt05:
    type: "cluster"
    collection: pflow
    dr: 0.5
    algorithm: antikt
    pt_min: 10
    nconst_min: 2
  ElectronIsolation:
    type: "isolation"
    collection: "electrons"
    dr: 0.4
  MuonIsolation:
    type: "isolation"
    collection: "muons"
    dr: 0.4

# -- Output --------------------------------------------
output:
  file_path: ""              # Output ROOT file path
  format: default
\end{lstlisting}

For the neural backend, the maximum supported particle count and the truth-level
$p_T$ selection cut are read from the released model metadata rather than set by the
user (e.g. a $0.25$\,GeV cut for the CMS 2011 model and $0.5$\,GeV for the ALEPH model).
For the released neural backend the pile-up treatment is model-defined:
the model trained on CMS 2011 pile-up conditions generates PFlow pile-up
particles implicitly from the learned detector response.

The parametric configuration additionally supports an optional
\texttt{pileup} block inside the generator section for Delphes-style pile-up
merging:

\begin{lstlisting}[style=yaml,caption={Parametric generator with optional
pile-up merging.},label=lst:pileup-config]
generator:
  type: "parametric"
  name: "cms"                    # "cms" or "atlas"
  seed: 42                       # random seed for reproducibility
  pileup:                        # optional pile-up configuration
    file_path: "MinBias.pileup"  # Delphes .pileup binary file
    mean_pileup: 50              # average PU interactions
    sigma_z: 0.053               # vertex z sigma (meters)
    sigma_t: 160e-12             # vertex t sigma (seconds)
    smear_hs_vertex: true        # also smear hard-scatter vertex
\end{lstlisting}

\begin{table}[htbp]
\centering
\caption{Dataset configuration fields.}
\label{tab:dataset-config}
\begin{tabular}{@{}lll p{0.6\textwidth}@{}}
\toprule
\textbf{Field} & \textbf{Type} & \textbf{Default} & \textbf{Description} \\
\midrule
\texttt{file\_path} & string & required & Input path, either a HepMC file or a Pythia8 \texttt{.cmnd} card. \\
\texttt{num\_events} & integer & \texttt{1} & Number of events to process. \\
\texttt{entry\_start} & integer & \texttt{0} & Starting entry for HepMC input; useful for splitting a file into non-overlapping chunks. \\
\bottomrule
\end{tabular}
\end{table}

\begin{table}[htbp]
\centering
\caption{Generator configuration fields.}
\label{tab:generator-config}
\begin{tabular}{@{}lll p{0.6\textwidth}@{}}
\toprule
\textbf{Field} & \textbf{Type} & \textbf{Default} & \textbf{Description} \\
\midrule
\texttt{type} & string & required & Backend type: \texttt{"neural"} or \texttt{"parametric"}. \\
\texttt{name} & string & required & Backend name from the registry, e.g. \texttt{"cms\_2011\_flow\_v00"}, \texttt{"aleph\_flow\_v00"}, \texttt{"cms"}, or \texttt{"atlas"}. \\
\texttt{batch\_size} & integer & \texttt{2000} & Number of events per generator batch. \\
\texttt{device} & string & \texttt{"cpu"} & Computation device for PyTorch-backed generators: \texttt{"cpu"}, \texttt{"cuda"}, or \texttt{"mps"}. \\
\texttt{num\_steps} & integer & \texttt{50} & Neural-only number of flow-matching integration steps. \\
\texttt{seed} & integer & none & Parametric-only random seed for reproducibility. \\
\texttt{debug} & boolean & \texttt{false} & Parametric-only option to write intermediate detector-stage collections. \\
\texttt{pileup} & object & none & Parametric-only Delphes-style pile-up merging block. \\
\bottomrule
\end{tabular}
\end{table}

\begin{table}[htbp]
\centering
\caption{Parametric pile-up fields.}
\label{tab:pileup-config}
\begin{tabular}{@{}lll p{0.6\textwidth}@{}}
\toprule
\textbf{Field} & \textbf{Type} & \textbf{Default} & \textbf{Description} \\
\midrule
\texttt{file\_path} & string & required & Path to a Delphes \texttt{.pileup} binary file. \\
\texttt{mean\_pileup} & float & required & Average number of pile-up interactions per bunch crossing. \\
\texttt{max\_z\_spread} & float & \texttt{0.25} & Vertex-$z$ truncation bound in meters. \\
\texttt{max\_t\_spread} & float & \texttt{800e-12} & Vertex-time truncation bound in seconds. \\
\texttt{sigma\_z} & float & \texttt{0.053} & Gaussian vertex-$z$ width in meters. \\
\texttt{sigma\_t} & float & \texttt{160e-12} & Gaussian vertex-time width in seconds. \\
\texttt{smear\_hs\_vertex} & boolean & \texttt{true} & Whether to smear the hard-scatter primary vertex. \\
\bottomrule
\end{tabular}
\end{table}

\begin{table}[htbp]
\centering
\caption{Common parameters to both \texttt{cluster} and \texttt{isolation} pipelines.}
\label{tab:pipeline-common}
\begin{tabularx}{\textwidth}{@{} ll l X @{}}
\toprule
\textbf{Parameter} & \textbf{Type} & \textbf{Default} & \textbf{Description} \\
\midrule
\texttt{batch\_size}    & integer & \texttt{2000} & Number of physics events managed per active post-processing chunk. \\
\texttt{num\_processes} & integer & \texttt{1}    & Total allocated worker processes. Setting to \texttt{1} forces synchronous runtime. \\
\bottomrule
\end{tabularx}
\end{table}

\begin{table}[htbp]
\centering
\caption{Configuration parameters for the jet clustering pipeline.}
\label{tab:pipeline-cluster}
\begin{tabularx}{\textwidth}{@{} ll l X @{}}
\toprule
\textbf{Parameter} & \textbf{Type} & \textbf{Default} & \textbf{Description} \\
\midrule
\texttt{type}            & string  & \textit{required} & Must be explicitly designated as \texttt{"cluster"}. \\
\texttt{collection}      & string  & \texttt{"pflow"}  & Baseline constituent particle collection to cluster: \texttt{"truth"} or \texttt{"pflow"}. \\
\texttt{algorithm}       & string  & \texttt{"antikt"} & Kinematic clustering metric selection: \texttt{"antikt"}, \texttt{"cambridge"}, \texttt{"genkt"}, or \texttt{"ee-genkt"}. \\
\texttt{dr}              & float   & \texttt{0.5}      & Spatial resolution jet radius parameter ($R$). \\
\texttt{algorithm\_param} & float  & \textit{none}     & Generalized power exponent $p$ for \texttt{"genkt"}/\texttt{"ee-genkt"} (where $p=1$ gives $k_t$, $p=0$ gives C/A, and $p=-1$ gives anti-$k_t$). \\
\texttt{pt\_min}          & float   & \texttt{0}        & Hard transverse momentum threshold ($p_{\text{T}}^{\text{min}}$) in GeV. \\
\texttt{nconst\_min}     & integer & \texttt{2}        & Lower boundary floor for required constituent multiplicity inside a jet. \\
\bottomrule
\end{tabularx}
\end{table}

\begin{table}[htbp]
\centering
\caption{Configuration parameters for the lepton isolation pipeline.}
\label{tab:pipeline-isolation}
\begin{tabularx}{\textwidth}{@{} ll l X @{}}
\toprule
\textbf{Parameter} & \textbf{Type} & \textbf{Default} & \textbf{Description} \\
\midrule
\texttt{type}       & string & \textit{required} & Must be explicitly designated as \texttt{"isolation"}. \\
\texttt{collection} & string & \texttt{"electrons"} & Base target object context: \texttt{"electrons"}, \texttt{"muons"}, or \texttt{"all"}. \\
\texttt{dr}         & float  & \texttt{0.4}      & Isolation cone spatial size constraint ($\Delta R$). \\
\bottomrule
\end{tabularx}
\end{table}

\begin{table}[htbp]
\centering
\caption{Configuration parameters for the filter pipeline.}
\label{tab:pipeline-filter}
\begin{tabularx}{\textwidth}{@{} ll l X @{}}
\toprule
\textbf{Parameter} & \textbf{Type} & \textbf{Default} & \textbf{Description} \\
\midrule
\texttt{type}           & string  & \textit{required} & Must be explicitly designated as \texttt{"filter"}. \\
\texttt{collection}     & string  & \texttt{"pflow"}  & Context string array key targeting a valid data structure (e.g., \texttt{"truth"}, \texttt{"Track"}). \\
\texttt{combine}        & string  & \texttt{"all"}    & Logical evaluation linkage rule: \texttt{"all"} ($\text{AND}$) or \texttt{"any"} ($\text{OR}$). \\
\texttt{conditions}     & list    & \texttt{[]}       & Array block specifying distinct cut dictionaries evaluated per entry (see block specifications below). \\
\texttt{update\_event\_features} & boolean & \texttt{true} & Flag forcing an automatic recalculation of event metrics ($H_{\text{T}}$ and $E_{\text{T}}^{\text{miss}}$) from surviving entries. \\
\midrule
\multicolumn{4}{@{}l}{\textbf{Structured Element Keys within the \texttt{conditions} Array Block}} \\
\midrule
\texttt{field} & string & \textit{required} & Particle or track attribute to query (e.g., \texttt{pt}, \texttt{eta}, \texttt{pdg\_id}, \texttt{et}). \\
\texttt{op}    & string & \textit{required} & Comparative constraint operator string chosen from: \texttt{>}, \texttt{>=}, \texttt{<}, \texttt{<=}, \texttt{==}, \texttt{!=}, \texttt{in}, \texttt{not in}. \\
\texttt{value} & \textit{varies} & \textit{required} & Target testing parameter value (scalar for comparisons, discrete array for set matches). \\
\texttt{abs}   & boolean & \texttt{false}    & Modifies evaluation parameter to process the absolute magnitude (e.g., $|\eta|$ evaluation). \\
\bottomrule
\end{tabularx}
\end{table}

\newpage

\section{TorchDelphes}\label{section:torch_delphes}

\subsection{Implementation} 

Delphes is composed of \textit{modules}, each of which drives a different part of the detector emulation model. 
For example, the \texttt{Efficiency} class takes a set of particles as input and randomly masks a specific fraction of them.
One could configure the \texttt{Efficiency} class to represent the efficiency with which CMS tracks charged hadrons, and call this specific configuration/instantiation the \texttt{CMSChargedHadronTrackingEfficiency} module.

To emulate an entire detector such as ATLAS or CMS, users configure modules to represent each subsystem
of their detector emulation model (e.g. charged hadron tracking, electron momentum smearing, calorimeter hit clustering),
and chain them together to define a computational graph for their intended simulation.
This is all done via a Tcl configuration file, several of which are shipped pre-built with the Delphes software. These include the default ATLAS configuration, \texttt{delphes\_card\_ATLAS.tcl}, and the default CMS configuration, \texttt{delphes\_card\_CMS.tcl}.

Parnassus.TorchDelphes defines PyTorch analogues of these C++ Delphes modules, providing both GPU acceleration and a Python interface.
Importantly, TorchDelphes does not define an analogue for \textit{every} Delphes module,
only the modules necessary up to the generation of calorimeter tower hits ("Towers") and reconstructed
photons, neutral hadrons, and charged tracks (collectively called "EnergyFlowObjects" or "EFlowObjects").
Later post-processing in a typical Delphes workflow, that is independent of the actual detector
interactions (e.g. jet clustering, object isolation), is handled by later \texttt{parnassus.pipelines} logic, not TorchDelphes.

In addition to these constituent modules, TorchDelphes provides three more classes,

\noindent\texttt{ATLASEnergyFlowDefault},
\texttt{CMSEnergyFlowDefault}, and \texttt{ALEPHEnergyFlowDefault}, which correspond to the default
ATLAS, CMS, and ALEPH configurations that ship with Delphes, respectively, again only up to the
creation of Towers and EFlowObjects.
These three default classes are used to run the validation suite.

\subsection{Validation}\label{section:torch_delphes_validation}

We validate the TorchDelphes implementation by comparing its intermediate and final outputs against benchmarks generated by C++ Delphes v3.5.1. We use the same HepMC inputs for both TorchDelphes and C++ Delphes, and these are generated by C++ Pythia8 v8.315, not Parnassus.Pythia. All of the code to run the full validation pipeline, including the configuration files to generate the HepMC inputs and the ROOT benchmarks, is available under the \texttt{torch\_delphes/validation} directory in the main github repository.

In this appendix we compare distributions generated by Parnassus.TorchDelphes against those 
generated by C++ Delphes for the same exact detector configurations and HepMC inputs.
We do this for the ATLAS, CMS, and ALEPH default detector configurations. For the
ATLAS and CMS cards we use two types of HepMC inputs,
$H \to Z Z \to 4 \ell$ and $t \bar{t} \to W$, while the ALEPH card is validated against
the $e^+e^- \to ZZ\to q\bar{q}$ process appropriate to its LEP environment.

For brevity, we limit this presentation to four particle-level kinematics variables - pseudorapidity ($\eta$), azimuthal angle ($\phi$), momentum/energy ($E/p$), and transverse momentum/energy ($E_T/p_T$). 
We also limit the scope to both \textit{final} output objects - \texttt{Tower} and \texttt{EFlowObject}, which are generated stochastically - and one \textit{intermediate} output object - \texttt{PropagatedParticle}, which is generated deterministically. 
However, users can validate the entire phase-space - that is, all of the particle-level distributions (e.g. charge, PDG ID) for all of the intermediate products (e.g. the outputs of the \texttt{SimpleCalorimeter} modules) - by passing the optional \texttt{--debug} parameter to the validation script\footnote{See the README in \texttt{torch\_delphes/validation}/}.
Users can also validate against individual events instead of large sets of events, results which are also presented here.

\begin{figure}[htbp]
\centering
\includegraphics[width=\textwidth]{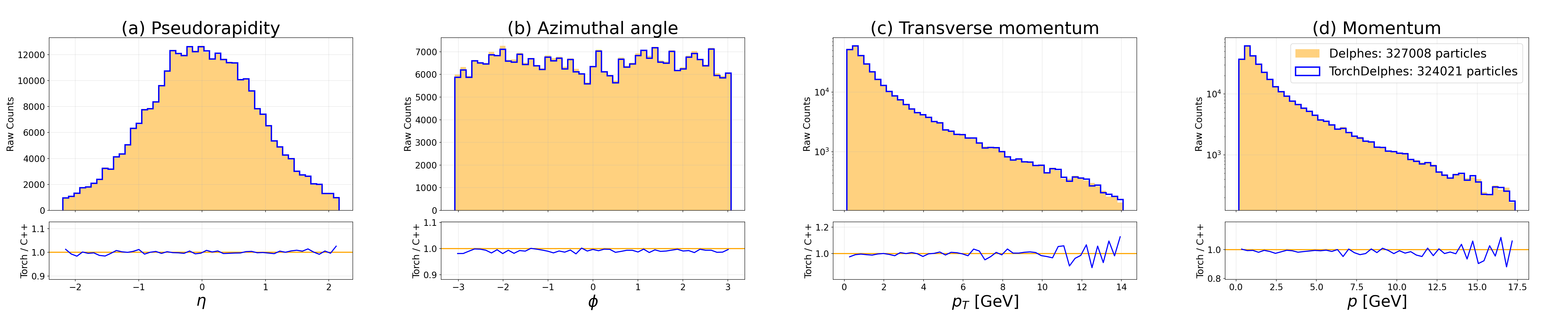}
\caption{Particle-level distributions comparing the \textsc{EFlowObject} outputs of \textsc{TorchDelphes} against those of \textsc{Delphes}. Both sets of outputs are generated via the default ALEPH detector configuration, from 10,000 truth-level $e^+e^- \to ZZ\to q\bar{q}$ \textsc{Pythia8} events. (a) Pseudorapidity, (b) azimuthal angle, (c) transverse momentum, (d) total momentum. Distributions are not normalized. } \label{fig:aleph_eflow}
\end{figure}

As expected, deterministic intermediate objects agree event-by-event between
\textsc{TorchDelphes} and C++ \textsc{Delphes}. 
Stochastic output objects are compared distributionally rather than event-by-event because the random streams
are not matched across the C++ and PyTorch implementations, and the two
programs may be run on different device types (e.g. CPU and GPU).
The same distributional agreement holds for the ALEPH card, whose
\textsc{EFlowObject} outputs are shown in~\autoref{fig:aleph_eflow}.

\begin{figure}[htbp]
\centering
\includegraphics[width=\textwidth]{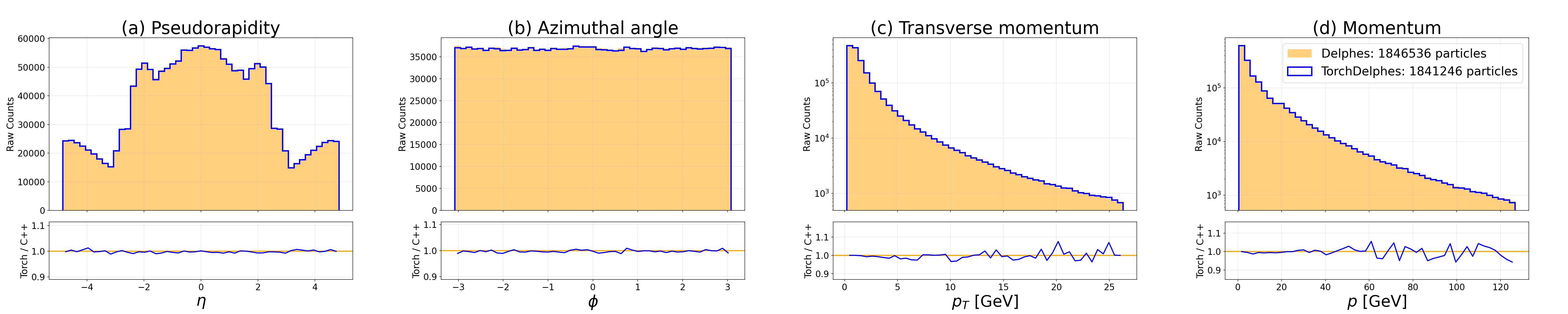}
\caption{Particle-level distributions comparing the \textsc{EFlowObject} outputs of \textsc{TorchDelphes} against those of \textsc{Delphes}. Both sets of outputs are generated via the default CMS detector configuration, from 10,000 truth-level $t\bar{t} \to W$ \textsc{Pythia8} events. (a) Pseudorapidity, (b) azimuthal angle, (c) transverse momentum, (d) total momentum. Distributions are not normalized.} \label{fig:cms_ttw_efobject}
\end{figure}

\begin{figure}[htbp]
\centering
\includegraphics[width=\textwidth]{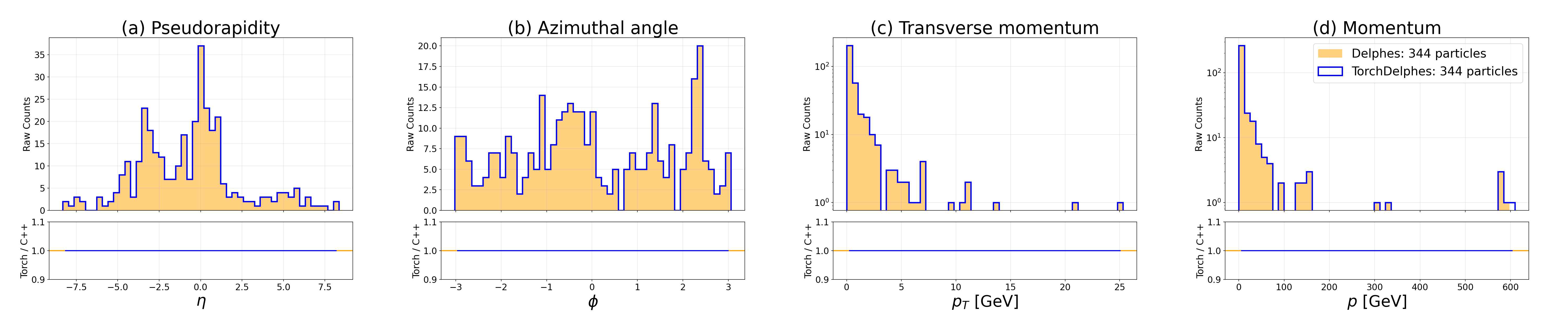}
\caption{Particle-level distributions comparing the \textsc{PropagatedParticle} outputs of \textsc{TorchDelphes} against those of \textsc{Delphes}, this time from a \textbf{single} truth-level $t\bar{t} \to W$ event, but still via the default CMS detector configuration. (a) Pseudorapidity, (b) azimuthal angle, (c) transverse momentum, (d) total momentum. Distributions are not normalized. These histograms show perfect agreement because the \textsc{PropagatedParticle} objects are deterministically generated.} \label{fig:cms_ttw_single_prop_part}
\end{figure}

\begin{figure}[htbp]
\centering
\includegraphics[width=\textwidth]{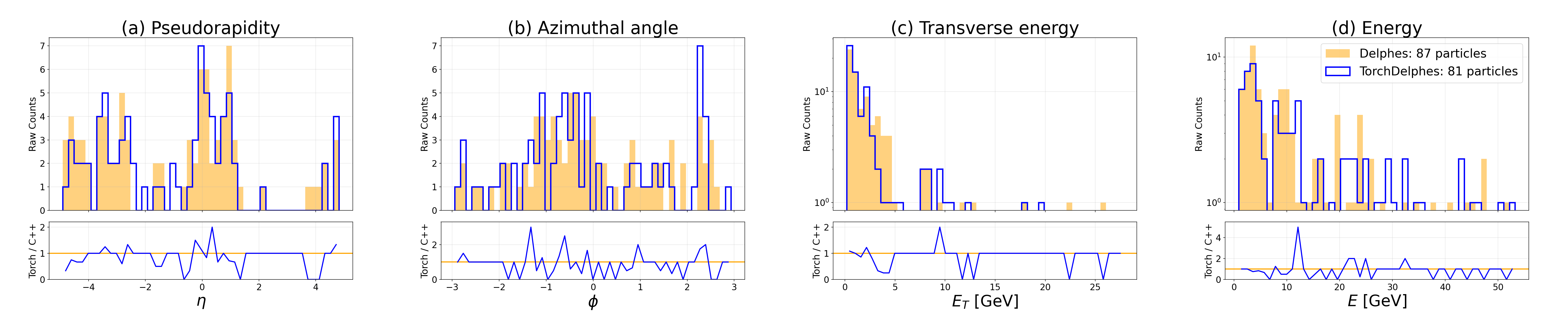}
\caption{Particle-level distributions comparing the \textsc{Tower} outputs of \textsc{TorchDelphes} against those of \textsc{Delphes}, from the same  truth-level $t\bar{t} \to W$ event and via the same default CMS detector configuration as~\autoref{fig:cms_ttw_single_prop_part}. (a) Pseudorapidity, (b) azimuthal angle, (c) transverse energy, (d) total energy. Distributions are not normalized. These histograms show imperfect agreement because the \textsc{Tower} objects are probabilistically/stochastically generated.} \label{fig:cms_ttw_single_tower}
\end{figure}

\begin{figure}[htbp]
\centering
\includegraphics[width=\textwidth]{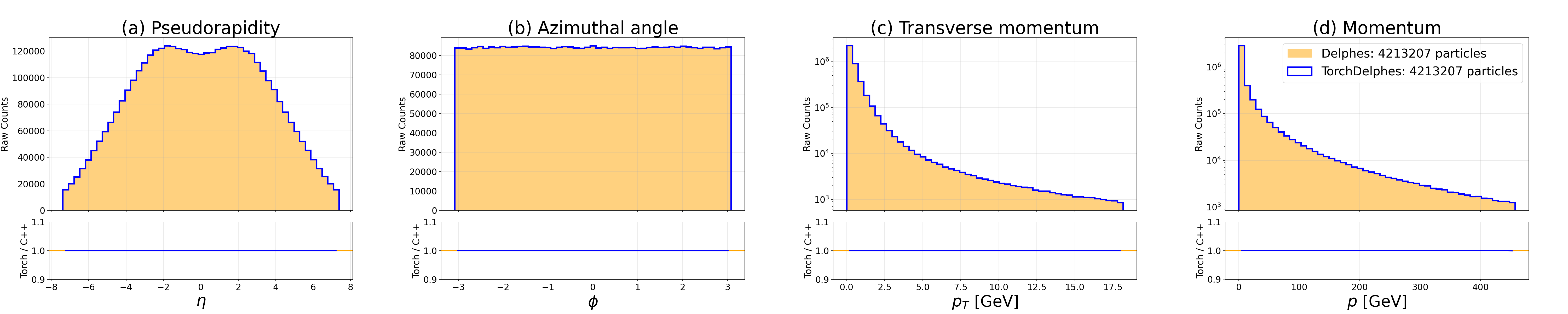}
\caption{Particle-level distributions comparing the \textsc{PropagatedParticle} outputs of \textsc{TorchDelphes} against those of \textsc{Delphes}, this time from 10,000 truth-level $H \to Z Z \to 4 \ell$ events and via the default ATLAS detector configuration. (a) Pseudorapidity, (b) azimuthal angle, (c) transverse momentum, (d) total momentum. Distributions are not normalized. Again, the deterministically-generated \textsc{PropagatedParticle} objects show perfect agreement between \textsc{TorchDelphes} and \textsc{Delphes}} \label{fig:atlas_hzz4l_part_prop}
\end{figure}

\begin{figure}[htbp]
\centering
\includegraphics[width=\textwidth]{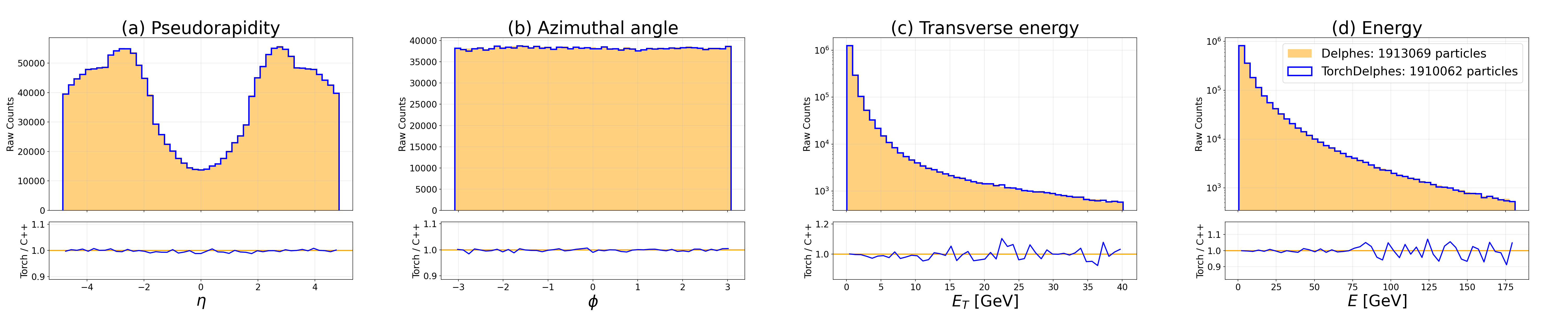}
\caption{Particle-level distributions comparing the \textsc{Tower} outputs of \textsc{TorchDelphes} against those of \textsc{Delphes},  from the same 10,000 truth-level $H \to Z Z \to 4 \ell$ events and via the same default ATLAS detector configuration as~\autoref{fig:atlas_hzz4l_part_prop}. (a) Pseudorapidity, (b) azimuthal angle, (c) transverse energy, (d) total energy. Distributions are not normalized. Again, the stochastically-generated \textsc{Tower} objects show imperfect agreement between \textsc{TorchDelphes} and \textsc{Delphes}, although the agreement is much better than that shown in~\autoref{fig:cms_ttw_single_tower} because of increased statistics. } \label{fig:atlas_hzz4l_tower}
\end{figure}

\newpage
\section{Input cards used in the demonstration}
\label{sec:demo-input-cards}

The following Pythia8 steering cards are the references used for the SM and BSM
examples in Sec.~\ref{sec:demo}.

\begin{lstlisting}[style=cmnd,caption={Pythia8 steering card for SM $H \to ZZ \to 4l$.},label=code:h4lep]
! h4lep.cmnd -- SM Higgs decay

Beams:idA = 2212                    ! first beam proton
Beams:idB = 2212                    ! second beam proton
Beams:eCM = 13000.                  ! beam COM energy

HiggsSM:gg2H = on                   ! Enable gluon fusion Higgs production
25:m0 = 125.0                       ! Set Higgs mass to 125 GeV
25:onMode = off
25:onIfMatch = 23 23                ! Force H->ZZ
23:onMode = off
23:onIfAny = 11 13                  ! Force Z->2l

! Particle decay limits (ct > 10 mm = stable)
ParticleDecays:limitTau0 = on
ParticleDecays:tau0Max = 10          ! mm/c
\end{lstlisting}

\begin{lstlisting}[style=cmnd,caption={Pythia8 steering card for BSM $H \to aa \to gg\mu\mu$.},label=code:bsm]
! haa_ggmumu.cmnd -- BSM exotic Higgs decay
Beams:eCM = 13000.
Beams:idA = 2212
Beams:idB = 2212

! BSM Higgs production via gluon fusion
Higgs:useBSM = on
HiggsBSM:gg2H2 = on
35:m0 = 125.0                       ! Higgs mass

! SM-like couplings for production
HiggsH2:coup2u = 1.0
HiggsH2:coup2d = 1.0
HiggsH2:coup2A3A3 = 1.0             ! H -> aa coupling

! Force H -> a a only
35:onMode = off
35:onIfAll = 36 36

! Light pseudoscalar a properties (PDG ID 36)
36:m0 = 20.0                        ! mass = 20 GeV
HiggsA3:coup2d = 1.0
HiggsA3:coup2u = 1.0
HiggsA3:coup2l = 1.0

! Force a -> gg or mumu only
36:onMode = off
36:onIfMatch = 21 21
36:onIfMatch = 13 -13 

! Particle decay limits (ct > 10 mm = stable)
ParticleDecays:limitTau0 = on
ParticleDecays:tau0Max = 10
\end{lstlisting}

\bibliographystyle{elsarticle-num}
\bibliography{references}

\begin{thebibliography}{10}
\expandafter\ifx\csname url\endcsname\relax
  \def\url#1{\texttt{#1}}\fi
\expandafter\ifx\csname urlprefix\endcsname\relax\def\urlprefix{URL }\fi
\expandafter\ifx\csname href\endcsname\relax
  \def\href#1#2{#2} \def\path#1{#1}\fi

\bibitem{delphes}
J.~de~Favereau, et~al., Delphes 3: A modular framework for fast simulation of a generic collider experiment, JHEP 02 (2014) 057.
\newblock \href {http://arxiv.org/abs/1307.6346} {\path{arXiv:1307.6346}}.

\bibitem{parnassus_v2}
E.~Dreyer, E.~Gross, D.~Kobylianskii, V.~Mikuni, B.~Nachman, Conditional deep generative models for simultaneous simulation and reconstruction of entire events (2025).
\newblock \href {http://arxiv.org/abs/2503.19981} {\path{arXiv:2503.19981}}.

\bibitem{parnassus_lep}
Y.~Lo, D.~Kobylianskii, B.~Nachman, An ai-based detector simulation and reconstruction model for the aleph experiment at lep (2026).
\newblock \href {http://arxiv.org/abs/2604.11834} {\path{arXiv:2604.11834}}.

\bibitem{fastjet}
M.~Cacciari, G.~P. Salam, G.~Soyez, Fastjet user manual, European Physical Journal C 72 (2012) 1896.
\newblock \href {http://arxiv.org/abs/1111.6097} {\path{arXiv:1111.6097}}.

\bibitem{uproot}
J.~Pivarski, et~al., \href{https://github.com/scikit-hep/uproot5}{Uproot: Root i/o in pure python and numpy}.
\newline\urlprefix\url{https://github.com/scikit-hep/uproot5}

\bibitem{geant4}
S.~Agostinelli, et~al., Geant4---a simulation toolkit, Nuclear Instruments and Methods in Physics Research Section A 506 (2003) 250.

\bibitem{lipman2023flow}
Y.~Lipman, R.~T.~Q. Chen, H.~Ben-Hamu, M.~Nickel, Flow matching for generative modeling, in: International Conference on Learning Representations, 2023.
\newblock \href {http://arxiv.org/abs/2210.02747} {\path{arXiv:2210.02747}}.

\bibitem{liu2023flow}
X.~Liu, C.~Gong, Q.~Liu, Flow straight and fast: Learning to generate and transfer data with rectified flow, in: International Conference on Learning Representations, 2023.
\newblock \href {http://arxiv.org/abs/2209.03003} {\path{arXiv:2209.03003}}.

\bibitem{root}
R.~Brun, F.~Rademakers, Root---an object oriented data analysis framework, Nuclear Instruments and Methods in Physics Research Section A 389 (1997) 81--86.
\newblock \href {https://doi.org/10.1016/S0168-9002(97)00048-X} {\path{doi:10.1016/S0168-9002(97)00048-X}}.

\bibitem{pytorch}
A.~Paszke, et~al., Pytorch: An imperative style, high-performance deep learning library, in: Advances in Neural Information Processing Systems, 2019.
\newblock \href {http://arxiv.org/abs/1912.01703} {\path{arXiv:1912.01703}}.

\bibitem{parnassus_v1}
E.~Dreyer, E.~Gross, D.~Kobylianskii, V.~Mikuni, B.~Nachman, N.~Soybelman, Parnassus: An automated approach to accurate, precise, and fast detector simulation and reconstruction (2024).
\newblock \href {http://arxiv.org/abs/2406.01620} {\path{arXiv:2406.01620}}.

\bibitem{buckleyHepMC3EventRecord2021}
A.~Buckley, P.~Ilten, D.~Konstantinov, L.~Lönnblad, J.~Monk, W.~Pokorski, T.~Przedzinski, A.~Verbytskyi, \href{http://arxiv.org/abs/1912.08005}{The {HepMC3} {Event} {Record} {Library} for {Monte} {Carlo} {Event} {Generators}}, Computer Physics Communications 260 (2021) 107310, arXiv:1912.08005 [hep-ph].
\newblock \href {https://doi.org/10.1016/j.cpc.2020.107310} {\path{doi:10.1016/j.cpc.2020.107310}}.
\newline\urlprefix\url{http://arxiv.org/abs/1912.08005}

\bibitem{bierlich_comprehensive_2022}
C.~Bierlich, S.~Chakraborty, N.~Desai, L.~Gellersen, I.~Helenius, P.~Ilten, L.~Lönnblad, S.~Mrenna, S.~Prestel, C.~T. Preuss, T.~Sjöstrand, P.~Skands, M.~Utheim, R.~Verheyen, \href{https://scipost.org/SciPostPhysCodeb.8}{A comprehensive guide to the physics and usage of {PYTHIA} 8.3}, SciPost Physics Codebases (2022) 008\href {https://doi.org/10.21468/SciPostPhysCodeb.8} {\path{doi:10.21468/SciPostPhysCodeb.8}}.
\newline\urlprefix\url{https://scipost.org/SciPostPhysCodeb.8}

\bibitem{ATLAS_SIMU_2026_09}
{ATLAS Collaboration}, \href{https://atlas.web.cern.ch/Atlas/GROUPS/PHYSICS/PLOTS/SIMU-2026-09/}{Combined fast simulation and reconstruction in {ATLAS} with {PARNASSUS}}, {ATLAS Public Plot PLOT-SIMU-2026-09}, accessed: 2026-06-23 (2026).
\newline\urlprefix\url{https://atlas.web.cern.ch/Atlas/GROUPS/PHYSICS/PLOTS/SIMU-2026-09/}

\end{thebibliography}

\end{document}